\documentclass[twocolumn,preprintnumbers,showpacs,amsmath,amssymb,prb]{revtex4}

\usepackage{graphicx}
\usepackage{dcolumn}
\usepackage{bm}
\usepackage{amssymb}
\usepackage{epstopdf}
\usepackage{color}

\begin{document}

\title{All-optical band engineering of gapped Dirac materials}
\author{O. V. Kibis$^{1,2,3}$}\email{Oleg.Kibis(c)nstu.ru}
\author{K. Dini$^2$}
\author{I. V. Iorsh$^{3,4}$}
\author{I. A. Shelykh$^{2,4}$}

\affiliation{${^1}$Department of Applied and Theoretical Physics,
Novosibirsk State Technical University, Karl Marx Avenue 20,
Novosibirsk 630073, Russia} \affiliation{$^2$Science Institute,
University of Iceland, Dunhagi 3, IS-107, Reykjavik, Iceland}
\affiliation{$^3$Division of Physics and Applied Physics, Nanyang
Technological University 637371, Singapore} \affiliation{$^4$ITMO
University, Saint Petersburg 197101, Russia}

\begin{abstract}
We demonstrate theoretically that the interaction of electrons in
gapped Dirac materials (gapped graphene and transition-metal
dichalchogenide monolayers) with a strong off-resonant
electromagnetic field (dressing field) substantially renormalizes
the band gaps and the spin-orbit splitting. Moreover, the
renormalized electronic parameters drastically depend on the field
polarization. Namely, a linearly polarized dressing field always
decreases the band gap (and, particularly, can turn the gap into
zero), whereas a circularly polarized field breaks the equivalence
of valleys in different points of the Brillouin zone and can both
increase and decrease corresponding band gaps. As a consequence,
the dressing field can serve as an effective tool to control spin
and valley properties of the materials and be potentially
exploited in optoelectronic applications.
\end{abstract}
\pacs{73.22.Pr, 78.67.Wj}

\maketitle

\section{Introduction}
Advances in laser physics and microwave technique achieved in
recent decades have made possible the use of high-frequency fields
as tools of flexible control of various atomic and
condensed-matter structures (so called ``Floquet engineering''
based on the Floquet theory of periodically driven quantum
systems~\cite{Hanggi_98,Kohler_05,GoldMan2014,Holthaus_16,Meinert_16}).
As a consequence, the properties of electronic systems driven by
oscillating fields are actively studied to exploit unique features
of composite states of field and matter. Particularly, electron
strongly coupled to electromagnetic field
--- also known as ``electron dressed by field'' (dressed
electron) --- has become a commonly used model in modern
physics~\cite{Scully_01,Cohen-Tannoudji_04}. Recently, the
physical properties of dressed electrons were studied in various
nanostructures, including quantum
wells~\cite{Wagner_10,Teich_13,Kibis_14,Morina_15,Pervishko_15,Dini_16},
quantum
rings~\cite{Kibis_13,Sigurdsson_14,Joibari_14,Koshelev_15},
graphene~\cite{Oka_09,Kibis_10,Kibis_11_1,Syzranov_13,Perez_14,Glazov_14,Kibis_16,Kristinsson_16},
and topological
insulators~\cite{Ezawa_2013,Usaj_14,Torres_14,Calvo_15,Mikami_16}.
Developing this excited scientific trend in the present article,
we elaborated the theory of dressed electrons for gapped Dirac
materials.

The discovery of graphene --- a monolayer of carbon atoms with
linear (Dirac) dispersion of
electrons~\cite{Novoselov2004,CastroNeto2009,DasSarma2011} ---
initiated studies of the new class of artificial nanostructures
known as Dirac materials. While graphene by itself is
characterized by the gapless electron energy spectrum, many
efforts have been dedicated towards fabrication Dirac materials
with the band gap between the valence and conduction bands (gapped
Dirac materials). The electron energy spectrum of the materials is
parabolic near band edges but turns into the linear Dirac
dispersion if the band gap vanishes. Therefore, electronic
properties of gapped Dirac materials substantially depend on the
value of the gap and, consequently, are perspective for
nanoelectronic
applications~\cite{Ferrari2014,Lensky2015,Novoselov2016}. Although
dressed condensed-matter structures are in focus of attention for
a long time, a consistent quantum theory of the gapped Dirac
materials strongly coupled to light was not elaborated before.
Since the electronic structure of Dirac materials differs
crucially from conventional condensed-matter structures, the known
theory of light-matter coupling cannot be directly applied to the
gapped Dirac materials. Moreover, it should be noted that gapped
Dirac materials are currently considered as a basis for new
generation of optoelectronic devices. Therefore, their optical
properties deserve special consideration. This motivated us to
fill this gap in the theory. To solve this problem in the present
study, we will focus on the two gapped Dirac materials pictured
schematically in Fig.~1. First of them is the graphene layer grown
on a hexagonal boron nitride substrate~\cite{Sachs2011,Jung2014},
where the band gap can be tuned in the broad range with an
external gate voltage~\cite{Kindermann2012} (see Fig.~1a). The
second is a transition metal dichalchogenide (TMDC) which is a
monolayer of atomically thin semiconductor of the type
$\mathrm{M}\mathrm{X}_2$, where $\mathrm{M}$ is a transition metal
atom ($\mathrm{Mo}$, $\mathrm{W}$, etc.) and $\mathrm{X}$ is a
chalcogen atom ($\mathrm{S}$, $\mathrm{Se}$, or
$\mathrm{Te}$)~\cite{Wang2012, Butler2013} (see Fig.~1b). The
specific feature of the TMDC compounds is the giant spin-orbit
coupling~\cite{Kosmider2013,Kormanyos_15} which is attractive for
using in novel spintronic and valleytronic devices~\cite{Mak2014}.
\begin{figure}[!h]
\includegraphics[width=1.0\columnwidth]{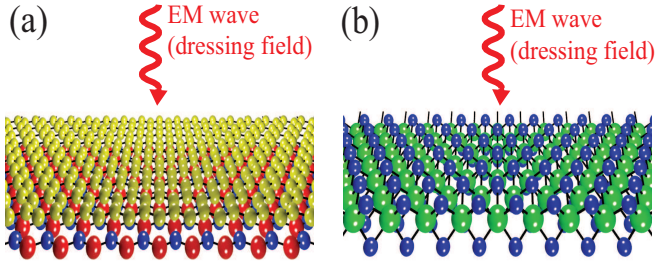}
\caption{(Color online) Sketch of the considered gapped Dirac
materials subject to electromagnetic wave (dressing field): (a)
Graphene grown on the substrate of hexagonal boron nitride; (b)
Transition metal dichalcogenide monolayer
$\mathrm{MoS}_2$.}\label{Fig.1}
\end{figure}
Formally, electronic properties of these materials near the band
edge can be described by the same two-band Hamiltonian
\begin{equation}\label{AH000}
\hat{{\cal H}}=\begin{pmatrix} \varepsilon_{\tau s}^c &
\gamma(\tau k_x- i k_y) \\ \gamma(\tau k_x +i k_y) &
\varepsilon_{\tau s}^v
\end{pmatrix},
\end{equation}
where $\mathbf{k}=(k_x,k_y)$ is the electron wave vector in the
layer plane, $\gamma$ is the parameter describing electron
dispersion,
\begin{equation}\label{Ec}
\varepsilon_{\tau s}^c=\frac{\Delta_g}{2}+\frac{\tau s
\Delta^{c}_{so}}{2}
\end{equation}
is the energy of the conduction band edge,
\begin{equation}\label{Ev}
\varepsilon_{\tau s}^v=-\frac{\Delta_g}{2}-\frac{\tau
s\Delta^v_{so}}{2}
\end{equation}
is the energy of the valence band edge, $\Delta_g$ is the band gap
between the conduction band and the valence band,
$\Delta_{so}^{c,v}$ is the spin-orbit splitting of the conduction
(valence) band, $s=\pm1$ is the spin index describing the
different spin orientations, and $\tau=\pm1$ is the valley index
which corresponds to the two valleys in the different points of
the Brillouin zone (the $K$ and $K^\prime$ valleys in
graphene~\cite{CastroNeto2009} and the $K$ and $-K$ valleys in
TMDC monolayers~\cite{Kormanyos_15}). If $\Delta_g\neq0$ and
$\Delta_{so}^{c,v}\neq0$, the Hamiltonian (\ref{AH0}) describes
TMDC monolayer \cite{Kormanyos_15}. In the case of zero spin-orbit
splitting, $\Delta_{so}^{c,v}=0$, the Hamiltonian (\ref{AH0})
describes gapped graphene~\cite{Sachs2011}, whereas the case of
$\Delta_g=\Delta_{so}^{c,v}=0$ corresponds to usual gapless
graphene~\cite{CastroNeto2009}. It should be noted that the
two-band Hamiltonian (\ref{AH000}) describes successfully
low-energy electron states near the band edge. As to omitted terms
corresponding to the trigonal-warping deformation of electron
bands in monolayer graphene, they can be neglected if the Rashba
spin-orbit coupling is stronger than the intrinsic spin-orbit
coupling~\cite{Rakyta_10}. In the present paper, we elaborate the
theory of electromagnetic dressing for electronic systems
described by the low-energy Hamiltonian (\ref{AH000}) and
demonstrate that both the band gap and the spin splitting can be
effectively controlled with the dressing field.

The paper is organized as follows. In the Section II, we apply the
conventional Floquet theory to derive the effective Hamiltonian
describing stationary properties of dressed electrons. In the
Section III, we discuss the dependence of renormalized electronic
characteristics of the dressed materials on parameters of the
dressing field. The last two Sections contain conclusion and
acknowledgements.

\section{Model}
Let us consider a gapped Dirac material with the Hamiltonian
(\ref{AH000}), which lies in the plane $(x,y)$ at $z=0$ and is
subjected to an electromagnetic wave propagating along the $z$
axis (see Fig.~1). The frequency of the wave, $\omega$, is assumed
to be far from all resonant frequencies of the electron system.
Therefore, the electromagnetic wave cannot be absorbed by
electrons near band edge and should be considered as a dressing
field for the states around $\textbf{k}=0$. Considering the
electron-field interaction within the minimal coupling approach,
properties of dressed electrons can be described by the
Hamiltonian
\begin{eqnarray}\label{AH0}
\hat{{\cal H}}(\mathbf{k})&=&\begin{pmatrix}0 & |e|\gamma(\tau
A_x-iA_y)/\hbar
\\|e|\gamma(\tau A_x+iA_y)/\hbar & 0\end{pmatrix}\nonumber\\
&+&\begin{pmatrix} \varepsilon_{\tau s}^c
& \gamma(\tau k_x- i k_y) \\
\gamma(\tau k_x +i k_y) & \varepsilon_{\tau s}^v
\end{pmatrix},
\end{eqnarray}
which can be easily obtained from the Hamiltonian of ``bare''
electrons (\ref{AH000}) with the replacement
$\mathbf{k}\rightarrow\mathbf{k}-(e/\hbar)\mathbf{A}$, where
$\mathbf{A}=(A_x,A_y)$ is the vector potential of the dressing
field, and $e$ is the electron charge. It should be noted that the
quantum electrodynamics predicts the quadratic (in the vector
potential) additions to the Hamiltonian
(\ref{AH0})~\cite{Chirolli_2012,Pellegrino_2014}. To avoid
complication of the model, we will assume that the considered
dressing field is classically strong and can be described
successfully with the minimal coupling. In what follows, we will
show that the properties of dressed electrons strongly depend on
the polarization of the dressing field. Therefore, we have to
discuss the solution of the corresponding Schr\"odinger problem
for different polarizations successively.

{\it Linearly polarized dressing field.---} Assuming the dressing
field to be linearly polarized along the $x$ axis, the vector
potential can be written as
\begin{equation}\label{AAl}
\mathbf{A}=\left(\frac{E_0}{\omega}\cos\omega t,0\right),
\end{equation}
where $E_0$ is the electric field amplitude, and $\omega$ is the
wave frequency. Correspondingly, the Hamiltonian of the dressed
electron system (\ref{AH0}) can be rewritten formally as
\begin{equation}\label{AHomega2}
\hat{\cal{H}}(\mathbf{k})=\hat{\cal{H}}_0+\hat{\cal{H}}_\mathbf{k},
\end{equation}
where
\begin{equation}\label{AH002}
\hat{\cal{H}}_0=\begin{pmatrix}
0 & \Omega\tau\hbar\omega/2\\
\Omega\tau\hbar\omega/2 & 0
\end{pmatrix}\cos\omega t
\end{equation}
is the Hamiltonian of electron-field interaction,
\begin{equation}\label{AHk2}
\hat{\cal{H}}_\mathbf{k}=
\begin{pmatrix}
\Delta_g/2+\tau s \Delta^{c}_{so}/2 & \gamma(\tau k_x-ik_y)\\
\gamma(\tau k_x+ik_y) & -\Delta_g/2-\tau s\Delta^v_{so}/2
\end{pmatrix}
\end{equation}
is the Hamiltonian of ``bare'' electron, and
\begin{equation}\label{Omega}
\Omega=\frac{2\gamma |e| E_0}{(\hbar \omega)^2}
\end{equation}
is the dimensionless parameter describing the strength of electron
coupling to the dressing field. The nonstationary Schr\"odinger
equation with the Hamiltonian (\ref{AH002}),
\begin{equation}\label{Ashr12}
i\hbar\frac{\partial\psi_0}{\partial t}=\hat{\cal{H}}_0\psi_0,
\end{equation}
describes the time evolution of electron states at the band edge
($\mathbf{k}=0$). The two exact solutions of the Schr\"odinger
problem (\ref{Ashr12}) read as
\begin{equation}\label{Apsi2}
\psi_0^\pm=\frac{1}{\sqrt{2}}
\begin{pmatrix}
\phantom{\pm}1\\
\pm1
\end{pmatrix}\exp\left[\mp\frac{i\Omega\tau\sin\omega
t}{2}\right].
\end{equation}
Since the two wave functions (\ref{Apsi2}) form a complete basis
at any fixed time $t$, we can seek solutions of the nonstationary
Schr\"odinger equation with the full Hamiltonian (\ref{AHomega2})
as an expansion
\begin{equation}\label{Apsik}
\psi_{\mathbf{k}}=a_1(t)\psi^+_0+a_2(t)\psi^-_0.
\end{equation}
Substituting the expansion (\ref{Apsik}) into the Schr\"odinger
equation,
\begin{equation}\label{Ashr112}
i\hbar\frac{\partial\psi_{\mathbf{k}}}{\partial
t}=\hat{\cal{H}}(\mathbf{k})\psi_{\mathbf{k}},
\end{equation}
we arrive at the expressions
\begin{eqnarray}\label{Aa}
i\hbar\dot{a}_1(t)&=&\left[\frac{\varepsilon_{\tau
s}^c+\varepsilon_{\tau s}^v}{2}+\gamma\tau
k_x\right]a_1(t)\nonumber\\
&+&\left[\frac{\varepsilon_{\tau s}^c-\varepsilon_{\tau
s}^v}{2}+i\gamma
k_y\right]e^{i\Omega\tau\sin\omega t}a_2(t),\nonumber\\
i\hbar\dot{a}_2(t)&=&\left[\frac{\varepsilon_{\tau
s}^c+\varepsilon_{\tau s}^v}{2}-\gamma\tau
k_x\right]a_2(t)\nonumber\\
&+&\left[\frac{\varepsilon_{\tau s}^c-\varepsilon_{\tau
s}^v}{2}-i\gamma k_y\right]e^{-i\Omega\tau\sin\omega t}a_1(t).
\end{eqnarray}
It follows from the conventional Floquet theory of quantum systems
driven by an oscillating
field~\cite{Hanggi_98,Kohler_05,GoldMan2014}  that the sought wave
function (\ref{Apsik}) must have the form
$\Psi(\mathbf{r},t)=e^{-i\tilde{\varepsilon}(\mathbf{k})
t/\hbar}\phi(\mathbf{r},t)$, where the function
$\phi(\mathbf{r},t)$ periodically depends on time,
$\phi(\mathbf{r},t)=\phi(\mathbf{r},t+2\pi/\omega)$, and
$\tilde{\varepsilon}(\mathbf{k})$ is the quasi-energy of an
electron. Since the quasi-energy (the energy of dressed electron)
is the physical quantity which plays the same role in quantum
systems driven by an oscillating field as the usual energy in
stationary ones, the present analysis of the Schr\"odinger problem
(\ref{Ashr112}) should be aimed to find the energy spectrum,
$\tilde{\varepsilon}(\mathbf{k})$. It follows from the periodicity
of the function $\phi(\mathbf{r},t)$ that one can seek the
coefficients $a_{1,2}(t)$ in Eq.~(\ref{Apsik}) as a Fourier
expansion,
\begin{equation}\label{AaF}
a_{1,2}(t)=e^{-i\tilde{\varepsilon}(\mathbf{k})
t/\hbar}\sum_{n=-\infty}^{\infty}a^{(n)}_{1,2} e^{in\omega t}.
\end{equation}
Substituting the expansion Eq.~\eqref{AaF} into the
Eqs.~\eqref{Aa} and applying the Jacobi-Anger expansion,
$$e^{iz\sin\theta}=\sum_{n=-\infty}^{\infty}J_n(z)e^{in\theta},$$
one can rewrite the equations of quantum dynamics (\ref{Aa}) in
the time-independent form,
\begin{equation}\label{Hk}
\sum_{n^\prime=-\infty}^{\infty}\sum_{j=1}^2{\cal
H}^{(nn^\prime)}_{ij}a^{(n^\prime)}_{j}=
\tilde{\varepsilon}(\mathbf{k})a^{(n)}_i,
\end{equation}
where $J_n(z)$ is the Bessel function of the first kind, and
${\cal H}^{(nn^\prime)}_{ij}$ is the stationary Hamiltonian of
dressed electron in the Floquet space with the matrix elements
\begin{eqnarray}\label{Hk0}
{\cal H}^{(nn^\prime)}_{12}&=&\left[\frac{\varepsilon_{\tau
s}^c-\varepsilon_{\tau s}^v}{2}+i\gamma
k_y\right]J_{n^\prime-n}\left(\Omega\tau\right),\nonumber\\
{\cal H}^{(nn^\prime)}_{21}&=&\left[\frac{\varepsilon_{\tau
s}^c-\varepsilon_{\tau s}^v}{2}-i\gamma
k_y\right]J_{n^\prime-n}\left(\Omega\tau\right),\nonumber\\
{\cal H}^{(nn^\prime)}_{11}&=&\left[\frac{\varepsilon_{\tau
s}^c+\varepsilon_{\tau s}^v}{2}+\gamma\tau
k_x+n\hbar\omega\right]\delta_{nn^{\prime}},\nonumber\\
{\cal H}^{(nn^\prime)}_{22}&=&\left[\frac{\varepsilon_{\tau
s}^c+\varepsilon_{\tau s}^v}{2}-\gamma\tau
k_x+n\hbar\omega\right]\delta_{nn^{\prime}},
\end{eqnarray}
where $\delta_{nn^{\prime}}$ is the Kronecker delta. It should be
noted that the Schr\"odinger equation (\ref{Hk}) describes still
exactly the initial Schr\"odinger problem (\ref{Ashr112}). Next we
will make some approximations.

In what follows, let us assume that the field frequency, $\omega$,
is high enough to satisfy the condition
\begin{equation}\label{hf}
\left|\frac{{\cal H}^{(0n)}_{ij}}{{\cal H}^{(00)}_{ii}-{\cal
H}^{(nn)}_{jj}}\right|\ll1
\end{equation}
for $n\neq0$ and $i\neq j$. Mathematically, the condition
(\ref{hf}) makes it possible to treat nondiagonal matrix elements
of the Hamiltonian (\ref{Hk0}) with $n\neq n^\prime$ as a small
perturbation which can be omitted in the first-order approximation
of the conventional perturbation theory for matrix Hamiltonians
(see, e.g., Ref.~\onlinecite{Bir1974}). Since the condition
(\ref{hf}) corresponds to an off-resonant field, the field can be
neither absorbed nor emitted by the electrons. As a consequence,
the main contribution to the Schr\"odinger equation (\ref{Hk})
under the condition (\ref{hf}) stems from terms with
$n,n^\prime=0$, which describe the elastic interaction between an
electron and the field. Neglecting the small terms with
$n,n^\prime\neq0$, the Schr\"odinger equation (\ref{Hk}) can be
rewritten in the simple form
\begin{equation}\label{Hk00}
\sum_{j=1}^2{\cal H}^{(00)}_{ij}a^{(0)}_{j}=
\tilde{\varepsilon}(\mathbf{k})a^{(0)}_i,
\end{equation}
where ${\cal H}^{(00)}_{ij}$ is the $2\times2$ Hamiltonian with
the matrix elements (\ref{Hk0}). Subjecting this Hamiltonian to
the unitary transformation
\begin{equation}\label{U}
U=\frac{1}{\sqrt{2}}\begin{pmatrix}
1&\phantom{-}1\\
1&-1\end{pmatrix},\nonumber
\end{equation}
we arrive at the effective stationary Hamiltonian of electrons
dressed by a linearly polarized field, $\hat{\cal
H}_{\mathrm{eff}}=U^\dagger{\cal H}^{(00)}U$, which is given by
the matrix
\begin{equation}\label{AHeff}
\hat{{\cal H}}_{\mathrm{eff}}(\mathbf{k})=\begin{pmatrix}
\widetilde{\Delta}_g/2+\tau s \widetilde{\Delta}^{c}_{so}/2 & \tau
\tilde{\gamma}_xk_x- i\tilde{\gamma}_y k_y \\
\tau\tilde{\gamma}_xk_x +i \tilde{\gamma}_yk_y &
-\widetilde{\Delta}_g/2-\tau s\widetilde{\Delta}^v_{so}/2
\end{pmatrix},
\end{equation}
where
\begin{align}\label{UP1}
\widetilde{\Delta}_g=\Delta_gJ_0(\Omega)
\end{align}
is the effective band gap,
\begin{align}\label{UPC2}
\widetilde{\Delta}^{c}_{so}=\frac{\Delta^c_{so}-\Delta^v_{so}}{2}+
\frac{\Delta^c_{so}+\Delta^v_{so}}{2}J_0(\Omega)
\end{align}
is the effective spin splitting of the conduction band,
\begin{align}\label{UPV2}
\widetilde{\Delta}^{v}_{so}=\frac{\Delta^v_{so}-\Delta^c_{so}}{2}+
\frac{\Delta^c_{so}+\Delta^v_{so}}{2}J_0(\Omega)
\end{align}
is the effective spin splitting of the valence band, and
\begin{align}\label{UP4}
\tilde{\gamma}_x= \gamma,\quad\tilde{\gamma}_y=\gamma J_0(\Omega)
\end{align}
are the effective parameters of electron dispersion along the
$x,y$ axes. The eigenenergy of the effective Hamiltonian
(\ref{AHeff}),
\begin{eqnarray}\label{Eeff}
\tilde{\varepsilon}^\pm_{\tau s}(\mathbf{k})&=&\frac{\tau
s(\widetilde{\Delta}_{so}^c-\widetilde{\Delta}_{so}^v)}{4}\\
&\pm&\sqrt{\left[ \frac{2\widetilde{\Delta}_g+\tau
s(\widetilde{\Delta}_{so}^c+\widetilde{\Delta}_{so}^v)}{4}\right]^2+\tilde{\gamma}_x^2\nonumber
k_x^2+\tilde{\gamma}_y^2 k_y^2},
\end{eqnarray}
is the sought energy spectrum of electrons dressed by the linearly
polarized field. Mathematically, the unperturbed Hamiltonian
(\ref{AH000}) is equal to the effective Hamiltonian (\ref{AHeff})
with the formal replacements
$\Delta_g\rightarrow\widetilde{\Delta}_g$,
$\Delta_{so}^{c,v}\rightarrow\widetilde{\Delta}_{so}^{c,v}$,
${\gamma}_{x,y}\rightarrow\tilde{\gamma}_{x,y}$. Therefore, the
behavior of a dressed electron is similar to the behavior of a
``bare'' electron with the renormalized band parameters
(\ref{UP1})--(\ref{UP4}). It should be noted that the effective
Hamiltonian (\ref{AHeff}) is derived under the condition
(\ref{hf}). Taking into account Eqs.~(\ref{Hk0}), the condition
(\ref{hf}) can be rewritten as $\gamma k, \Delta_g\ll\hbar\omega$.
Therefore, the effective Hamiltonian is applicable to describe the
dynamics of dressed electron near the band edge if the photon
energy of the dressed field, $\hbar\omega$, substantially exceeds
the band gap, $\Delta_g$. It should be noted that such high-energy
photons will lead to direct electron transitions between the
valence band and the conduction band. However, the transitions
take place very far from the band edge and do not effect on the
considered renormalization of low-energy electron states.

{\it Circularly polarized dressing field.---} For the case of
circularly polarized electromagnetic wave, the vector potential
$\mathbf{A}=(A_x,A_y)$ can be written as
\begin{equation}\label{AAc}
\mathbf{A}=\left(\frac{E_0}{\omega}\cos\xi\omega
t,\frac{E_0}{\omega}\sin\xi\omega t\right),
\end{equation}
where the different chirality indices $\xi=\pm1$ correspond to the
clockwice/counterclockwise circular polarizations. First of all,
let us consider electron states at the wave vector $\mathbf{k}=0$,
where the Hamiltonian (\ref{AH0}) can be written in the form
\begin{equation}\label{AH00}
\hat{\cal{H}}(0)=
\begin{pmatrix}
\varepsilon_{\tau s}^c & -(\hbar\omega\Omega\tau/2) e^{-i\tau\xi\omega t}\\
-(\hbar\omega\Omega\tau/2) e^{i\tau\xi\omega t} &
\varepsilon_{\tau s}^v
\end{pmatrix},
\end{equation}
which is similar to the well-known Hamiltonian of magnetic
resonance. The corresponding nonstationary Schr\"odinger equation,
\begin{equation}\label{Ashr1}
i\hbar\frac{\partial\psi_{\tau s}(0)}{\partial
t}=\hat{\cal{H}}(0)\psi_{\tau s}(0),
\end{equation}
describes the time evolution of electron states at the wave vector
$\mathbf{k}=0$. Solutions of the equation (\ref{Ashr1}) can be
sought as
\begin{equation}\label{Apsi1}
\psi^\pm_{\tau s}(0)=e^{-i\tilde{\varepsilon}^\pm_{\tau s}(0) t/\hbar}\begin{pmatrix} A_\pm e^{-i\tau\xi\omega t/2}\\
B_\pm e^{i\tau\xi\omega t/2}
\end{pmatrix}e^{\pm i\tau\xi\omega t/2},
\end{equation}
where $\tilde{\varepsilon}^\pm_{\tau s}(0)$, $A_\pm$ and $B_\pm$
are the undefined constants. Substituting the wave function
(\ref{Apsi1}) into the Schr\"odinger equation (\ref{Ashr1}) with
the Hamiltonian (\ref{AH00}), we arrive at the system of two
algebraic equations,
\begin{eqnarray}\label{AAB1}
A_\pm\left[\varepsilon_{\tau
s}^c-\tau\xi\frac{\hbar\omega}{2}(1\mp1)-\tilde{\varepsilon}^\pm_{\tau s}(0)\right]-B_\pm\frac{\hbar\omega\Omega\tau}{2}=0,\nonumber\\
A_\pm\frac{\hbar\omega\Omega\tau}{2}-B_\pm\left[\varepsilon_{\tau
s}^v+\tau\xi\frac{\hbar\omega}{2}(1\pm1)-\tilde{\varepsilon}^\pm_{\tau
s}(0)\right]=0,
\end{eqnarray}
which can be easily solved. As a result, the two orthonormal exact
solutions of the Schr\"odinger problem (\ref{Ashr1}) are
\begin{eqnarray}\label{Apsi21}
\psi_{\tau s}^{\pm}(0)&=&e^{-i\tilde{\varepsilon}^\pm_{\tau s}(0)
t/\hbar}e^{\pm
i\tau\xi\omega t/2}\nonumber\\
&\times&\begin{pmatrix}
\mp\left[\frac{\sqrt{\Omega^2+\delta^2}\pm|\delta|}{2\sqrt{\Omega^2+\delta^2}}\right]^{1/2}
e^{-i\tau\xi\omega t/2}\\
\mathrm{sgn}(\delta)\left[\frac{\sqrt{\Omega^2+\delta^2}\mp|\delta|}{2\sqrt{\Omega^2+\delta^2}}\right]^{1/2}e^{i\tau\xi\omega
t/2}\end{pmatrix},
\end{eqnarray}
where
\begin{equation}\label{Ene}
\tilde{\varepsilon}^\pm_{\tau s}(0)=\frac{\varepsilon_{\tau
s}^c+\varepsilon_{\tau
s}^v}{2}\pm\tau\xi\frac{\hbar\omega}{2}\pm\mathrm{sgn}(\delta)\frac{\hbar\omega}{2}
\sqrt{\Omega^2+\delta^2}
\end{equation}
is the quasienergy (energy of dressed electron in the
conduction/valence band) at $\mathbf{k}=0$, and
$$\delta=\frac{\varepsilon_{\tau s}^c-\varepsilon_{\tau s}^v-\tau\xi\hbar\omega}{\hbar\omega}$$
is the resonance detuning assumed to be nonzero in order to avoid
the field absorption near the band edge. Correspondingly, the
effective stationary Hamiltonian of dressed electron states at
$\mathbf{k}=0$ can be written in the basis (\ref{Apsi21}) as
\begin{equation}\label{AHeff1}
\hat{{\cal H}}_{\mathrm{eff}}(0)=\begin{pmatrix}
\tilde{\varepsilon}^+_{\tau s}(0) & 0 \\
0 & \tilde{\varepsilon}^-_{\tau s}(0)
\end{pmatrix}.
\end{equation}
In order to find the energy spectrum of dressed electron at the
wave vector $\mathbf{k}\neq0$, let us restrict the consideration
by the case of $\Omega\ll1$, which corresponds physically to high
frequencies $\omega$ [see Eq.~(\ref{Omega})]. Expanding the
electron wave function, $\psi_{\tau s}(\mathbf{k})$, on the basis
(\ref{Apsi21}),
\begin{equation}\label{Psi2}
\psi_{\tau s}(\mathbf{k})=a^+(t)e^{i\tilde{\varepsilon}^+_{\tau
s}(0) t/\hbar}\psi_{\tau
s}^+(0)+a^-(t)e^{i\tilde{\varepsilon}^-_{\tau s}(0)
t/\hbar}\psi_{\tau s}^-(0),
\end{equation}
and substituting the expansion (\ref{Psi2}) into the Schr\"odinger
equation with the total Hamiltonian (\ref{AH0}), we arrive at the
system of equations
\begin{align}\label{aa}
&i\hbar\dot{a}^+(t)\approx\tilde{\varepsilon}^+_{\tau
s}(0)a^+(t)-\mathrm{sgn}(\delta)\gamma(\tau
k_x-ik_y)a^-(t),\nonumber\\
&i\hbar\dot{a}^-(t)\approx\tilde{\varepsilon}^-_{\tau
s}(0)a^-(t)-\mathrm{sgn}(\delta)\gamma(\tau k_x+ik_y)a^+(t).
\end{align}
The quantum dynamics equations (\ref{aa}) are equal to the
stationary Schr\"odinger equation,
$$i\hbar \frac{\partial }{\partial
t}\begin{pmatrix}a^+(t)\\a^-(t)\end{pmatrix}=\hat{{\cal
H}}_{\mathrm{eff}}(\mathbf{k})\begin{pmatrix}a^+(t)\\a^-(t)\end{pmatrix},$$
where
\begin{equation}\label{AHeffk}
\hat{{\cal H}}_{\mathrm{eff}}(\mathbf{k})=\begin{pmatrix}
\tilde{\varepsilon}^+_{\tau s}(0) &
-\mathrm{sgn}(\delta)\gamma(\tau
k_x-ik_y) \\
-\mathrm{sgn}(\delta)\gamma(\tau k_x+ik_y) &
\tilde{\varepsilon}^-_{\tau s}(0)
\end{pmatrix}.
\end{equation}
is the effective stationary Hamiltonian of the considered system.
The eigenenergy of the Hamiltonian,
\begin{equation}\label{Eee}
\tilde{\varepsilon}_{\tau
s}^\pm(\mathbf{k})=\frac{\tilde{\varepsilon}_{\tau
s}^+(0)+\tilde{\varepsilon}_{\tau
s}^-(0)}{2}\pm\sqrt{\left[\frac{\tilde{\varepsilon}_{\tau
s}^+(0)-\tilde{\varepsilon}_{\tau s}^-(0)}{2}\right]^2+(\gamma
k)^2},
\end{equation}
presents the sought energy spectrum of dressed electrons. If
$\Delta_g=\Delta_{so}^{c,v}=0$, Eq.~(\ref{Eee}) exactly coincides
with the known spectrum of electrons in gapless graphene
irradiated by a circularly polarized light \cite{Kristinsson_16}.
It follows from Eq.~(\ref{Eee}) that the renormalized band gap is
\begin{align}\label{12a}
&\widetilde{\Delta}_{g}=\tau \xi\hbar \omega +
\mathrm{sgn}\left(\Delta_{g}-\tau \xi \hbar
\omega\right)\hbar\omega
\sqrt{\Omega^2 + \left[\frac{\Delta_{g}-\tau \xi \hbar \omega}{\hbar \omega}\right]^2}\nonumber\\
&\approx\tau \xi\hbar \omega - \sqrt{\Omega^2 + 1} \left( \tau \xi
\hbar \omega -\frac{\Delta_g }{\Omega^2+1}\right),
\end{align}
where the last equality holds under condition $\hbar \omega \gg
\Delta_g$. The spin splittings in the conduction and valence bands
can be written in simple form for the two limiting cases:
\begin{align}\label{hw_big_so}
\widetilde{\Delta}_{so}^{c,v}=\pm\frac{\Delta_{so}^c-\Delta_{so}^v}{2}+\frac{\Delta_{so}^c+\Delta_{so}^v}{2\sqrt{1+\Omega^2}},\,\,\,\,\,\hbar\omega\gg
\Delta_g,
\end{align}
and
\begin{align}\label{hw_sma_so}
&
\widetilde{\Delta}_{so}^{c,v}=\pm\frac{\Delta_{so}^c-\Delta_{so}^v}{2}+\frac{\Delta_{so}^c+\Delta_{so}^v}{2}\left[1-\frac{\Omega^2}{2}\frac{(\hbar\omega)^2}{\Delta_g^2}\right],\nonumber \\
&\hbar\omega\ll \Delta_g.
\end{align}
As expected, the renormalized band gap (\ref{12a}) and spin
splittings (\ref{hw_big_so})--(\ref{hw_sma_so}) turn into their
``bare'' values, ${\Delta}_{g}$ and ${\Delta}_{so}^{c,v}$, if the
dressing field is absent ($E_0\rightarrow0$).

{\it Elliptically polarized dressing field.---} Assuming the large
axis of polarization ellipse to be oriented along the $x$ axis,
the vector potential of arbitrary polarized electromagnetic wave,
$\mathbf{A}=(A_x,A_y)$, can be written as
\begin{equation}\label{AAA}
\mathbf{A}=\frac{E_0}{\omega} \Big(\cos\omega
t,\,\sin\theta\sin\omega t\Big),
\end{equation}
where $\theta\in[-\pi/2,\pi/2]$ is the polarization phase: the
polarization is linear for $\theta=0$, circular for
$\theta=\pm\pi/2$, and elliptical for other phases $\theta$.
Substituting the vector potential (\ref{AAA}) into
Eq.~(\ref{AH0}), we can write the total Hamiltonian (\ref{AH0}) as
\begin{align}
\hat{\mathcal{H}}(\mathbf{k})=\hat{\mathcal{H}}_{\mathbf{k}}+\left(\hat{V}e^{i\omega
t}+\hat{V}^\dagger e^{-i\omega t}\right),
\end{align}
where the Hamiltonian $\hat{\mathcal{H}}_{\mathbf{k}}$ is given by
Eq.~(\ref{AHk2}) and
\begin{align}\label{V}
\hat{V}=\frac{\hbar\omega\Omega}{4}\begin{pmatrix}0 &
\tau-\sin\theta
\\ \tau+\sin\theta& 0\end{pmatrix}.
\end{align}
is the operator of electron interaction with the dressing field
(\ref{AAA}). Generally, the effective stationary Hamiltonian of an
electron driven by an oscillating field can be sought in the
form~\cite{GoldMan2014}
\begin{align}\label{Hel}
\hat{{\cal
H}}_{\mathrm{eff}}(\mathbf{k})=e^{i\hat{F}(t)}\hat{\mathcal{H}}(\mathbf{k})e^{-i\hat{F}(t)}+i\left(\frac{\partial
e^{i\hat{F}(t)}}{\partial t}\right)e^{-i\hat{F}(t)},
\end{align}
where $\hat{F}(t)$ is the anti-Hermitian operator which is
periodical with the period of the oscillating field,
$\hat{F}(t)=\hat{F}(t+2\pi/\omega)$. In the particular case of
weak electron-field coupling, $\Omega\ll1$, this operator and the
effective Hamiltonian (\ref{Hel}) can be easily found as power
series expansions,
\begin{equation}\label{F}
\hat{F}(t)=\displaystyle\sum_{n=1}^{\infty}\frac{F^{(n)}(t)}{\omega^n},\,\,\,\,\,\,\,\,\,\,\,
\hat{\cal
H}_{\mathrm{eff}}(\mathbf{k})=\displaystyle\sum_{n=0}^{\infty}\frac{\hat{\cal
H}_{\mathrm{eff}}^{(n)}(\mathbf{k})}{\omega^n},
\end{equation}
where $F^{(n)}(t)\sim \Omega^n$ (the Floquet-Magnus
expansion~\cite{GoldMan2014}). Substituting the expansions
(\ref{F}) into Eq.~(\ref{Hel}) and restricting the accuracy by
terms $\sim\Omega^2$, we arrive at the effective Hamiltonian
\begin{equation}\label{Hel0}
\hat{\cal
H}_{\mathrm{eff}}(\mathbf{k})=\hat{\mathcal{H}}_{\mathbf{k}}+\frac{\left[\hat{V},\hat{V}^{\dagger}\right]}{\hbar\omega}+\frac{[[\hat{V},\hat{\mathcal{H}}_{\mathbf{k}}],\hat{V}^{\dagger}]+h.c.}{2(\hbar\omega)^2}.
\end{equation}
Taking into account Eqs.~(\ref{AHk2}) and (\ref{V}), the effective
stationary Hamiltonian (\ref{Hel0}) can be written as a matrix
(\ref{AHeff}), where
\begin{align}\label{11}
\widetilde{\Delta}_g=
\Delta_g\left[1-\frac{\Omega^2}{4}(1+\sin^2\theta)\right]
-\frac{\tau\hbar\omega\Omega^2}{2}\sin\theta,
\end{align}
\begin{eqnarray}\label{12}
\widetilde{\Delta}^{c,v}_{so}&=&\pm\frac{\Delta^c_{so}-\Delta^v_{so}}{2}+
\frac{\Delta^c_{so}+\Delta^v_{so}}{2}\nonumber\\
&\times&\left[1-\frac{\Omega^2}{4}\left(1+\sin^2\theta\right)\right],
\end{eqnarray}
\begin{align}\label{14}
\tilde{\gamma}_x=
\gamma\left[1-\frac{\Omega^2}{4}\sin^2\theta\right],\quad\tilde{\gamma}_y=\gamma\left[1-\frac{\Omega^2}{4}\right]
\end{align}
are the band parameters renormalized by an elliptically polarized
dressing field. Correspondingly, the eigenenergy of the effective
Hamiltonian (\ref{Hel0}) represents the sought energy spectrum of
dressed electrons,
\begin{eqnarray}\label{Eeffe}
\tilde{\varepsilon}^\pm_{\tau s}(\mathbf{k})&=&\frac{\tau
s(\widetilde{\Delta}_{so}^c-\widetilde{\Delta}_{so}^v)}{4}\\
&\pm&\sqrt{\left[ \frac{2\widetilde{\Delta}_g+\tau
s(\widetilde{\Delta}_{so}^c+\widetilde{\Delta}_{so}^v)}{4}\right]^2+\tilde{\gamma}_x^2\nonumber
k_x^2+\tilde{\gamma}_y^2 k_y^2},
\end{eqnarray}
with the renormalized band parameters (\ref{11})--(\ref{14}). It
should be stressed that the effective Hamiltonian (\ref{AHeff})
with the band parameters (\ref{11})--(\ref{14}), which describes
electrons dressed by an arbitrary polarized weak field, is derived
under assumption of small coupling constant (\ref{Omega}) and high
frequency, $\omega$. On the contrary, the effective Hamiltonian
(\ref{AHeff}) with the band parameters (\ref{UP1})--(\ref{UP4})
and the effective Hamiltonian (\ref{AHeff1}) are suitable to
describe electrons dressed by linearly and circularly polarized
dressing fields of arbitrary intensity. As a consequence, the band
parameters (\ref{UP1})--(\ref{UP4}) and
(\ref{12a})--(\ref{hw_big_so}) turn into the band parameters
(\ref{11})--(\ref{14}) for $\Omega\ll1$, $\hbar\omega\gg\Delta_g$,
and $\theta=0,\pm\pi/2$.

\section{Results and Discussion}

\begin{figure}[h]
\includegraphics[width=1.0\columnwidth]{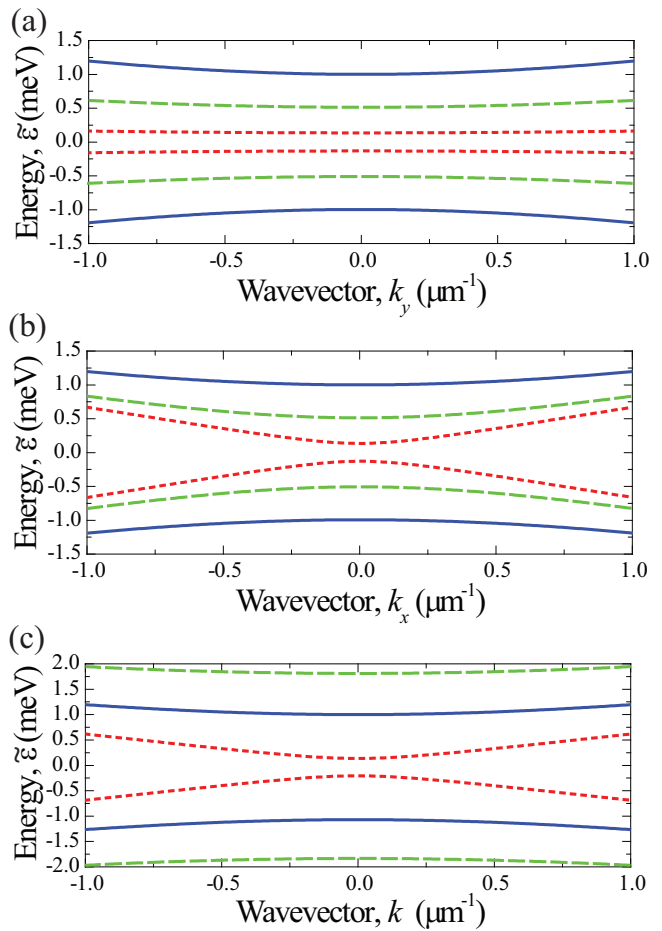}
\caption{(Color online) The energy spectrum of dressed electron,
$\tilde{\varepsilon}(\mathbf{k})$, near the band edge of gapped
graphene ($\Delta_g=2$~meV, $\gamma/\hbar=10^6$~m/s) irradiated by
a dressing field with the photon energy $\hbar\omega=10$~meV and
the different intensities, $I$. In the parts (a) and (b): the
dressing field is linearly polarized along the $x$ axis; the
irradiation intensities are $I=0$ (solid lines), $I=7.5$
kW$/$cm$^2$ (dashed lines), $I=15$ kW$/$cm$^2$ (dotted lines). In
the part (c): the dressing field is circularly polarized; the
solid line describes the energy spectrum of ``bare'' electron
($I=0$), whereas the dotted and dashed lines correspond to the
different circular polarizations ($\tau\xi=-1$ and $\tau\xi=1$,
respectively) with the same irradiation intensity $I=300$
W$/$cm$^2$.}\label{Fig.2}
\end{figure}
\begin{figure}[h]
\includegraphics[width=1.0\columnwidth]{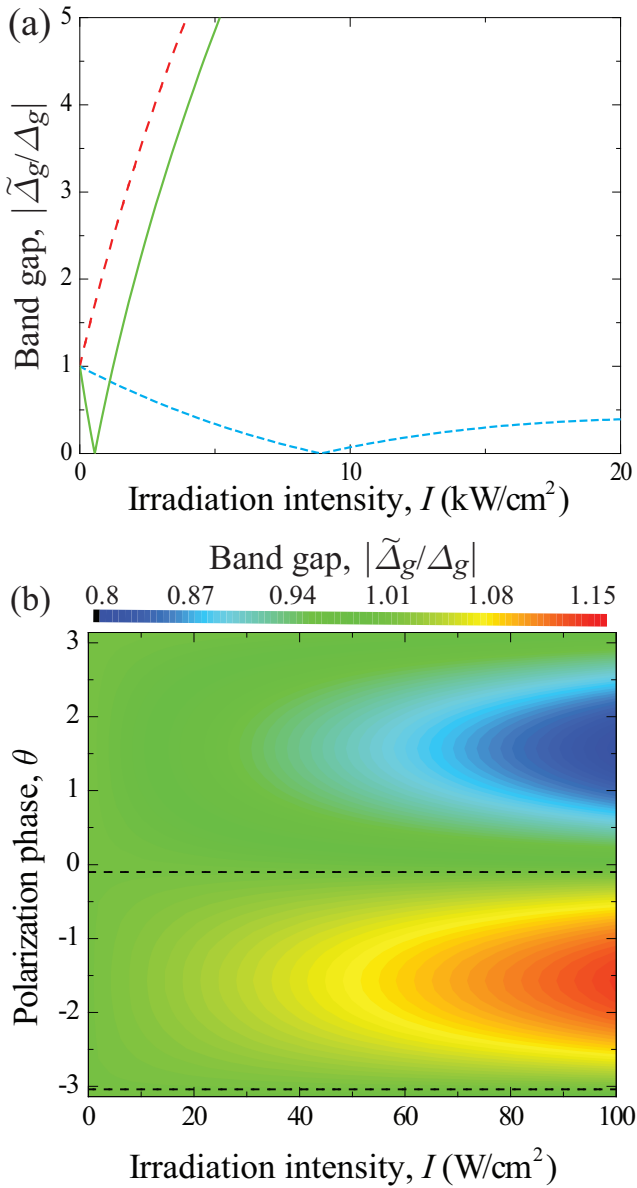}
\caption{(Color online) Dependence of the band gap in irradiated
gapped graphene ($\Delta_g=2$~meV, $\gamma/\hbar=10^6$~m/s) on the
irradiation intensity, $I$, and the polarization, $\theta$, for
the photon energy $\hbar\omega=10$~meV. In the part (a): the
dotted line corresponds to the linearly polarized dressing field,
whereas the dashed and solid lines correspond to the different
circular polarizations ($\tau\xi=-1$ and $\tau\xi=1$,
respectively). In the part (b): the dashed lines correspond to the
polarizations, $\theta$, which do not change the band
gap.}\label{Fig.3}
\end{figure}
First of all, let us apply the developed theory to gapped
graphene, assuming $\widetilde{\Delta}_{so}^{c,v}=0$ in all
derived expressions. The electron dispersion in gapped graphene,
$\tilde{\varepsilon}(\mathbf{k})$, is plotted in Fig.~2 for the
particular cases of linearly and circularly polarized dressing
field. It the absence of the dressing field, the electron
dispersion is isotropic in the graphene plane (see the solid lines
in Figs.~2a and 2b). However, a linearly polarized field breaks
the equivalence of the $x,y$ axes [see Eq.~(\ref{Eeff})]. As a
consequence, the anisotropy of the electron dispersion along the
wave vectors $k_x$ and $k_y$ appears (see the dashed and dotted
lines in Figs.~2a and 2b). In contrast to the linear polarization,
a circularly polarized dressing field does not induce the in-plane
anisotropy [see Eq.~(\ref{Eee})]. However, the electron dispersion
is substantially different for clockwise and counterclockwise
polarizations (see the dashed and dotted lines in Fig.~2c).
Moreover, both linearly and circularly polarized field
renormalizes the band gap (see Fig.~3). Mathematically, the
dependence of the renormalized band gap, $|\widetilde{\Delta}_g|$,
on the irradiation intensity, $I\sim E_0^2$, is given by
Eqs.~(\ref{UP1}) and (\ref{12a}) [which are plotted in Fig.~3a]
and Eq.~(\ref{11}) [which is plotted in Fig.~3b]. It should be
noted that Eq.~(\ref{12a}) correctly describes the gap for any
off-resonant frequencies $\omega$, whereas Eqs.~(\ref{UP1}) and
(\ref{11}) are derived under the condition
$\hbar\omega\gg\Delta_g$ and, therefore, applicable only to small
gaps. However, the gap can be gate-tunable in the broad range,
$\Delta_g=1-60$~meV~\cite{Sachs2011,Jung2014,Kindermann2012}.
Assuming the gap to be of meV scale and the field frequency to be
in the terahertz range, we can easily satisfy this condition. It
follows from Eqs.~(\ref{UP1}), (\ref{Ene}) and (\ref{11}) that the
renormalized gap, $\widetilde{\Delta}_g$, crucially depends on the
field polarization. Particularly, the clockwise/counterclockwise
circularly polarized field (polarization indices $\xi=\pm1$)
differently interacts with electrons from different valleys of the
Brillouin zone (valley indices $\tau=\pm1$). Namely, for the case
of $\tau\xi=-1$, the value of the gap monotonously increases with
intensity (see the dashed line in Fig.~3a). On the contrary, for
the case of $\tau\xi=1$, the gap first decreases to zero and then
starts to grow (see the solid line in Fig.~3a). This light-induced
difference in the band gaps for the two different valleys is
formally equivalent to the appearance of an effective magnetic
field acting on the valley pseudo-spin and, therefore, can be
potentially used in valleytronics applications. It should be noted
that this optically-induced lifting of valley degeneracy has been
observed for TMDC in the recent experiments~\cite{Sie_2015} which
are in reasonable agreement with the present theory. As to
linearly polarized dressing field, it always quenches the band gap
and can even turn it into zero (see the dotted line in Fig.~3a).
Formally, the collapse of the band gap originates from zeros of
the Bessel function in Eq.~(\ref{UP1}). Since the linearly and
circularly polarized fields change the gap value oppositely, there
are field polarizations which do not change the gap. The
polarization phases, $\theta$, corresponding to such polarizations
are marked by the dashed lines in Fig.~3b.

Applying the elaborated theory to analyze the renormalized spin
splitting in TMDC monolayers, let us restrict the consideration by
the most examined TMDC monolayer $\mathrm{MoS_2}$. The dependence
of the spin splitting on the dressing field is described by
Eqs.~(\ref{UPC2})--(\ref{UPV2}) for the case of linearly polarized
field, Eqs.~(\ref{hw_big_so})--(\ref{hw_sma_so}) for the case of
circularly polarized field, and Eq.~(\ref{12}) for an arbitrary
polarized field. It follows from analysis of these expressions
that the most pronounced renormalization of the splitting takes
place for a circularly polarized field. The dependence of the
field-induced renormalization of the band gap,
$\widetilde{\Delta}_{g}$, and the spin splitting,
$\widetilde{\Delta}_{so}^{c,v}$, in $\mathrm{MoS_2}$ monolayer on
the field intensity, $I\sim E_0^2$, is plotted in Fig.~4 for such
a field. It is seen in Fig.~4a that absolute values of the
field-induced renormalization are of the same order for both the
band gap, $\widetilde{\Delta}_{g}-{\Delta}_{g}$, and the spin
splitting, $\widetilde{\Delta}_{so}^{c,v}-{\Delta}_{so}^{c,v}$.
However, the unperturbed spin splitting of the conduction band,
${\Delta}_{so}^{c}$, is small as compared to both the unperturbed
band gap, ${\Delta}_{g}$, and the unperturbed spin splitting of
the valence band, ${\Delta}_{so}^{v}$ (see
Ref.~\onlinecite{Kormanyos_15}). Therefore, the relative
field-induced renormalization,
$|\widetilde{\Delta}_{so}^{c}/{\Delta}_{so}^{c}|$, is most
pronounced for the spin splitting of the conduction band (see
Fig.~4b). It should be stressed that the renormalized splitting
depends on the product of the polarization and valley indices,
$\tau\xi=\pm1$, and can be turned into zero by a dressing field
(see Fig.~4b). It should be noted also that the renormalization of
both band gap and spin splitting is the result of mixing electron
states from the valence and conduction bands by the field.
Therefore, the renormalized band parameters strongly depends on
the value of the ``bare'' band gap, ${\Delta}_g$. In contrast to
gapped graphene with band gaps of meV scale, TMDC monolayers have
band gaps of eV scale~\cite{Kormanyos_15}. As a consequence, the
considered terahertz photons effect on the gaps of TMDC very
weakly (in contrast to the previously considered case of
narrow-gapped graphene). Particularly, it follows from this that
the irradiation intensity which collapses the spin splitting in
TMDCs monolayers (see Fig.~4b) is really large than the intensity
collapsing the band gap in gapped graphene (see Fig.~3a).

It should be noted that the similar optically-induced spin
splitting was recently observed experimentally in
GaAs~\cite{Ryzhov_2016}. However, the one-band energy spectrum of
conduction electrons in GaAs differs crucially from the electron
spectrum of gapped Dirac materials describing by the two-band
Hamiltonian (\ref{AH000}). Therefore, the known theory of
optically-induced spin splitting for electrons with simple
parabolic dispersion --- including both the recent
paper~\cite{Ryzhov_2016} and the classical
article~\cite{Pershan_1966} --- cannot by applied directly to the
materials under consideration. One has to take also into account
that optical properties of TMDC are dominated by
excitons~\cite{He_2014,Chernikov_2014}. To avoid the influence of
excitons on the discussed dressing-field effects, the photon
energy, $\hbar\omega$, should be less than the binding exciton
energy (which is typically of hundreds of meV in TMDC).

\begin{figure}[h]
\includegraphics[width=1.0\columnwidth]{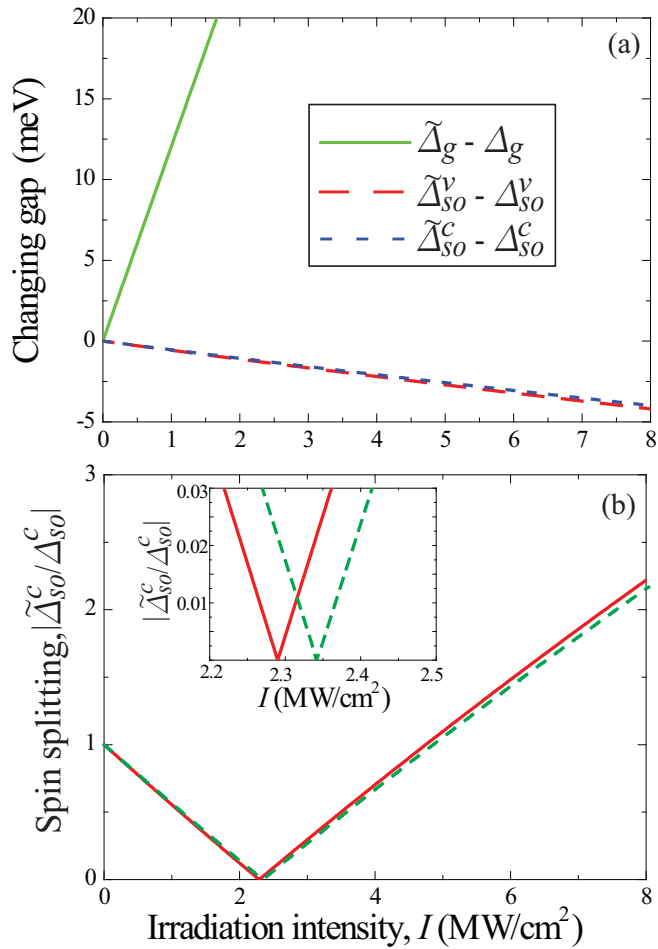}
\caption{ (Color online) Dependence of the band gap,
$\widetilde{\Delta}_{g}$, and the spin splitting of conduction and
valence bands, $\widetilde{\Delta}_{so}^{c,v}$, on the irradiation
intensity for $\mathrm{MoS}_2$ monolayer ($\Delta_g=1.58$ eV,
$\Delta_{so}^c = 3$ meV, $\Delta_{so}^v = 147$ meV,
$\gamma/\hbar=7.7 \times 10^5$ m/s) irradiated by a circularly
polarized field with the photon energy $\hbar\omega=10$~meV: (a)
The field-induced changing of the band gap and the spin splitting
for the circularly polarized field with $\tau\xi=-1$; (b) The
renormalized spin-slitting of conduction band for different field
polarizations (the solid and dashed lines correspond to
$\tau\xi=-1$ and $\tau\xi=1$, respectively).}\label{Fig.4}
\end{figure}

From viewpoint of experimental observability of the discussed
phenomena, it should be noted that all dressing-field effects
increase with increasing the intensity of the dressing field.
However, an intense irradiation can melt a condensed-matter
sample. To avoid the melting, it is reasonable to use narrow
pulses of a strong dressing field. This well-known methodology has
been elaborated long ago and commonly used to observe various
dressing effects --- particularly, modifications of energy
spectrum of dressed electrons arisen from the optical Stark effect
--- in semiconductor structures (see, e.g., Refs.~\onlinecite{Joffre_1988_1,Joffre_1988_2,Lee_1991}). Within this approach,
giant dressing fields (up to GW/cm$^2$) can be applied to the
structures. It should be noted also that we consider the
electromagnetic wave as a purely dressing field which cannot be
absorbed by electrons. Within the classical Drude theory, the
collisional absorption of the oscillating field (\ref{AAl}) by
conduction electrons is given by the well-known expression
\begin{equation}\label{Drude}\nonumber
Q=\frac{1}{T}\int_0^T
\mathbf{j}(t)\mathbf{E}(t)\mathrm{d}t=\frac{E_0^2}{2}\frac{\sigma_0}{1+(\omega\tau_0)^2},
\end{equation}
where $T$ is the period of the field, $Q$ is the period-averaged
field energy absorbed by conduction electrons per unit time and
per unit volume, $\mathbf{j}(t)$ is the ohmic current density
induced by the oscillating electric field
$\mathbf{E}(t)=E_0\sin\omega t$, $\sigma_0$ is the static Drude
conductivity, and $\tau_0$ is the electron relaxation time.
Evidently, the Drude optical absorption, $Q$, is negligibly small
under the condition $\omega\tau_0\gg1$. Thus, an electromagnetic
wave can be considered as a purely dressing field in the
high-frequency limit (see, e.g., Ref.~\onlinecite{Kibis_14} for
more details). It should be stressed that the increasing of
temperature decreases the time $\tau_0$ because of the
strengthening of the electron-phonon scattering. Therefore, the
temperature should be low enough to meet the aforementioned
condition.

\section{Conclusion}

We showed that the electromagnetic dressing can be used as an
effective tool to control various electronic properties of gapped
Dirac materials, including the band gap in gapped graphene and the
spin splitting in TMDC monolayers. Particularly, both the band gap
and the spin splitting can be closed by a dressing field. It is
demonstrated that the strong polarization dependence of the
renormalized band parameters appears. Namely, a linearly polarized
field decreases the band gap, whereas a circularly polarized field
can both decrease and increase one. It is found also that a
circularly polarized field breaks equivalence of valleys in
different points of the Brillouin zone, since the renormalized
band parameters depend on the valley index. As a consequence, the
elaborated theory creates a physical basis for novel electronic,
spintronic and valleytronic devices operated by light.

\begin{acknowledgments}
The work was partially supported by the RISE project CoExAN, FP7
ITN project NOTEDEV, RFBR projects 16-32-60123 and 17-02-00053,
the Rannis projects 141241-051 and 163082-051, and the Russian
Ministry of Education and Science (projects 3.1365.2017,
3.2614.2017 and 3.4573.2017). O.V.K. and I.V.I. acknowledge
support from the Singaporean Ministry of Education under AcRF Tier
2 grant MOE2015-T2-1-055.

\end{acknowledgments}

\end{document}